\documentclass[%
 superscriptaddress,
 preprint,
 showpacs,preprintnumbers,
 amsmath,amssymb,
 aps,
 prb,
]{revtex4-2}

\usepackage{graphicx}% Include figure files
\usepackage{dcolumn}% Align table columns on decimal point
\usepackage{bm}% bold math
\usepackage{upgreek}
\usepackage{amsmath}

\begin{document}

%\preprint{APS/123-QED}

\title{Enhanced sum-frequency generation from etchless lithium niobate empowered by dual quasi-bound states in the continuum}% Force line breaks with \\

\author{Siqi Feng}
\affiliation{Institute for Advanced Study, Nanchang University, Nanchang 330031, China}
\affiliation{Jiangxi Key Laboratory for Microscale Interdisciplinary Study, Nanchang University, Nanchang 330031, China}

\author{Tingting Liu}
\email{ttliu@ncu.edu.cn}
\affiliation{Institute for Advanced Study, Nanchang University, Nanchang 330031, China}
\affiliation{Jiangxi Key Laboratory for Microscale Interdisciplinary Study, Nanchang University, Nanchang 330031, China}

\author{Wenya Chen}
\affiliation{Institute for Advanced Study, Nanchang University, Nanchang 330031, China}
\affiliation{Jiangxi Key Laboratory for Microscale Interdisciplinary Study, Nanchang University, Nanchang 330031, China}

\author{Feng Wu}
\email{fengwu@gpnu.edu.cn}
\affiliation{School of Optoelectronic Engineering, Guangdong Polytechnic Normal University, Guangzhou 510665, China}

\author{Shuyuan Xiao}
\email{syxiao@ncu.edu.cn}
\affiliation{Institute for Advanced Study, Nanchang University, Nanchang 330031, China}
\affiliation{Jiangxi Key Laboratory for Microscale Interdisciplinary Study, Nanchang University, Nanchang 330031, China}

\begin{abstract}
	
The miniaturization of nonlinear light sources is central to the integrated photonic platform, driving a quest for high-efficiency frequency generation and mixing at the nanoscale. In this quest, the high-quality ($Q$) resonant dielectric nanostructures hold great promise, as they enhance nonlinear effects through the resonantly local electromagnetic fields overlapping the chosen nonlinear materials. Here, we propose a method for the enhanced sum-frequency generation (SFG) from etcheless lithium niobate (LiNbO$_{3}$) by utilizing the dual quasi-bound states in the continuum (quasi-BICs) in a one-dimensional resonant grating waveguide structure. Two high-$Q$ guided mode resonances corresponding to the dual quasi-BICs are respectively excited by two near-infrared input beams, generating a strong visible SFG signal with a remarkably high conversion efficiency of $3.66\times10^{-2}$ (five orders of magnitude higher than that of LiNbO$_{3}$ films of the same thickness) and a small full-width at half-maximum less than 0.2 nm. The SFG efficiency can be tuned via adjusting the grating geometry parameter or choosing the input beam polarization combination. Furthermore, the generated SFG signal can be maintained at a fixed wavelength without the appreciable loss of efficiency by selectively exciting the angle-dependent quasi-BICs, even if the wavelengths of input beams are tuned within a broad spectral range. Our results provide a simple but robust paradigm of high-efficiency frequency conversion on an easy-fabricated platform, which may find applications in nonlinear light sources and quantum photonics.

\end{abstract}

%\pacs{42.70.-a, 42.79.-e, 78.67.Pt}% PACS, the Physics and Astronomy
                             % Classification Scheme.
%\keywords{Suggested keywords}%Use showkeys class option if keyword
                              %display desired
\maketitle

%\tableofcontents

\section{\label{sec1}Introduction}

Frequency conversion via sum-frequency generation (SFG) is of paramount importance for a wide variety of applications such as coherent light sources, single photon detection, and vibrational hyperspectral imaging. It is a three-wave mixing process where two incident photons with frequencies $\omega_{1}$ and $\omega_{2}$ interact inside a second-order nonlinear material to generate coherent light with their sum frequency $\omega_{\text{SFG}}=\omega_{1}+\omega_{2}$\cite{Boyd2020}. The efficient SFG emission is generally realized using the bulk transparent nonlinear crystals and the intense pulsed laser beams in conjunction with phase-matching techniques. When the thickness of nonlinear materials becomes less than the input wavelength, the phase-matching is no longer applicable to improve the conversion efficiency. Recently, strong resonant modes in dielectric nanostructures, including nanoantennas and metasurfaces, have been proposed to boost the frequency conversion, via taking advantage of the complete overlap between the structures and the enhanced local electromagnetic fields through multipolar modes of Mie-type resonances\cite{Smirnova2016, Sain2019, Grinblat2021, Liu2022}. Compared to harmonic generations, the nondegenerate nature of SFG process involving with the input beams at different wavelengths requires the design of multi-resonant nanostructures\cite{Rocco2022}. Dual-resonance enhanced SFG has been reported in some pioneering works, underpinned by magnetic and electric multipolar resonances\cite{Liu2018, Frizyuk2019, Marino2019, Zilli2021, Nikolaeva2021, CamachoMorales2021}. In these studies, the quality ($Q$) factor related to the capability of local electric field confinement, becomes a principal indicator for the enhancement of nonlinear light-matter interactions, which leads to significant research interests in high-$Q$ resonances for high-efficiency SFG.

The ultrahigh-$Q$ resonances based on bound states in the continuum (BICs) have suggested a promising solution to facilitate light-matter interactions in dielectric nanostructures\cite{Zhen2014, Huang2020, Wang2020, Dong2022, Kang2022, Qin2022, Chen2023, Zhong2023}. BICs are the peculiar states embedded to the radiation continuum but remain perfectly confined without any radiation. A genuine BIC exists in ideal lossless infinite structures, with an infinitely high $Q$ factor and vanishing resonant width, and can be put into reality as a quasi-BIC by introducing perturbations in various forms\cite{Hsu2016, Sadreev2021, Huang2023}. In the past decade, the quasi-BICs with remarkable local electric field confinement have been practically utilized to boost the harmonic generations\cite{Carletti2018, Xu2019, Koshelev2019, Carletti2019, Liu2019, Koshelev2020, Anthur2020, Gandolfi2021, Zograf2022, Xiao2022, Xiao2022a, Zalogina2023}. The high-efficiency SFG can be envisaged in engineered nanostructures that support quasi-BICs at both input wavelengths. Very recently, this physical mechanism has been successfully demonstrated in symmetry-broken metasurfaces made of high-index nonlinear dielectric materials like III-V group semiconductors\cite{CamachoMorales2022}, and further applied to the spontaneous parametric down-conversion (SPDC), the reverse process of SFG\cite{Parry2021, Mazzanti2022, Santiago-Cruz2022, Son2023}. However, it may fail if the high-index materials were replaced with the low-index, such as lithium niobate (LiNbO$_{3}$). The quasi-BICs would suffer from a strong leakage to the surroundings, because of the low index contrast between LiNbO$_{3}$ ($n_{\text{e}}\approx2.15$, $n_{\text{0}}\approx2.22$ at 1.55 $\upmu$m\cite{Zelmon1997}) and the regular supporting substrate (usually SiO$_{2}$ with $n\approx1.45$). Moreover, the high cost has long been a challenge to the large-scale etching process of patterning subwavelength resonators out of LiNbO$_{3}$ material\cite{Timpu2019, Li2020, Fedotova2020, Yuan2021, Zhang2022, Ma2021, Qu2022}. Despite the tremendous advantages of LiNbO$_{3}$ over III-V group semiconductors in particular the broadband transparency window from ultraviolet to infrared\cite{Fedotova2022, Boes2023}, the above-mentioned factors significantly impede the development of LiNbO$_{3}$-based SFG process.

In this work, we present a simple and robust approach for the high-efficiency SFG from etchless LiNbO$_{3}$, by fully exploiting the dual quasi-BICs in a one-dimensional resonant grating waveguide structure which is composed of a four-part periodic grating and a LiNbO$_{3}$ waveguide layer. In the design, two guided mode resonances corresponding to the dual quasi-BICs can be respectively excited at the wavelengths of the two input beams in the infrared regime, generating a strong visible SFG signal. Benefiting from their ultrahigh-$Q$ factors, the SFG conversion efficiency is substantially increased by five orders of magnitude compared with the LiNbO$_{3}$ films of the same thickness, and in addition, the SFG photons are emitted only into a narrow linewidth with full-width at half-maximum less than 0.2 nm. By adjusting the grating geometry parameter, the SFG efficiency can be tuned over two orders of magnitude arising from the varying $Q$ factors; owing to the polarization sensitivity of the one-dimensional structure, it can also be continuously tuned between its maximum and minimum via simply choosing the input beam polarization combination. More interesting, the narrowband SFG signal can be generated at a fixed wavelength for the input beams within a broad wavelength range without the appreciable loss of efficiency, by tuning spectral locations of the angle-dependent quasi-BICs. These results suggest a way for smart engineering of the quasi-BICs in dielectric nanostructures for nonlinear devices and applications.

\section{\label{sec2}Dual quasi-bound states in the continuum} 

The schematic of the SFG process is shown in Fig. 1(a). Two pump photons with frequencies $\omega_{1}$ and $\omega_{2}$ impinge from the grating side, generate a photon with their sum frequency $\omega_{\text{SFG}}=\omega_{1}+\omega_{2}$ in the waveguide layer, and finally the SFG signal gets emitted from the substrate side. The proposed resonant grating waveguide structure is designed to support dual quasi-BICs at $\omega_{1}$ and $\omega_{2}$, respectively. It consists of a four-part periodic arrangement of grating on top of a LiNbO$_{3}$ waveguide layer residing on substrate. To support the guided mode, the effective refractive indices of the grating and substrate should be lower than that of the LiNbO$_{3}$ waveguide layer. Without loss of generality, we choose SiO$_{2}$ as the constituent material here since it has been widely adopted in the most versatile LiNbO$_{3}$-on-insulator platform for both linear and nonlinear applications\cite{Wang2017, Yu2020, Chen2021, Ye2022, Zhang2022a, Huang2022, Li2022, Chen2023b}. Figure 1(b) shows the unit cell of the structure. For the four-part periodic grating, the first and third parts made of SiO$_{2}$ share the same width $d_{\text{a}}$, while the second and fourth parts filled with air differ as $d_{\text{b}}=d_{0}-\Delta d$ and $d_{\text{c}}=d_{0}+\Delta d$, respectively, where $d_{0}$ is their initial width. The ratio between $\Delta d$ and $d_{0}$, i.e., $\alpha=\Delta d/d_{0}$, is defined as the grating geometry parameter. The width difference between the second and fourth parts within the unit cell ($0\textless\alpha\textless1$) would introduce perturbations into the grating waveguide structure, and two ultrahigh-$Q$ guided mode resonances arise as a result of the transition from the genuine BICs to the quasi-BICs, affording the resonantly local electric field confinement for the nonlinear SFG process. The heights of grating and waveguide are $h_{\text{G}}$ and $h_{\text{WG}}$, respectively, and the substrate is assumed to be semi-infinitely thick.

\begin{figure*}[htbp]
	\centering
	\includegraphics% Here is how to import EPS art
	[scale=0.36]{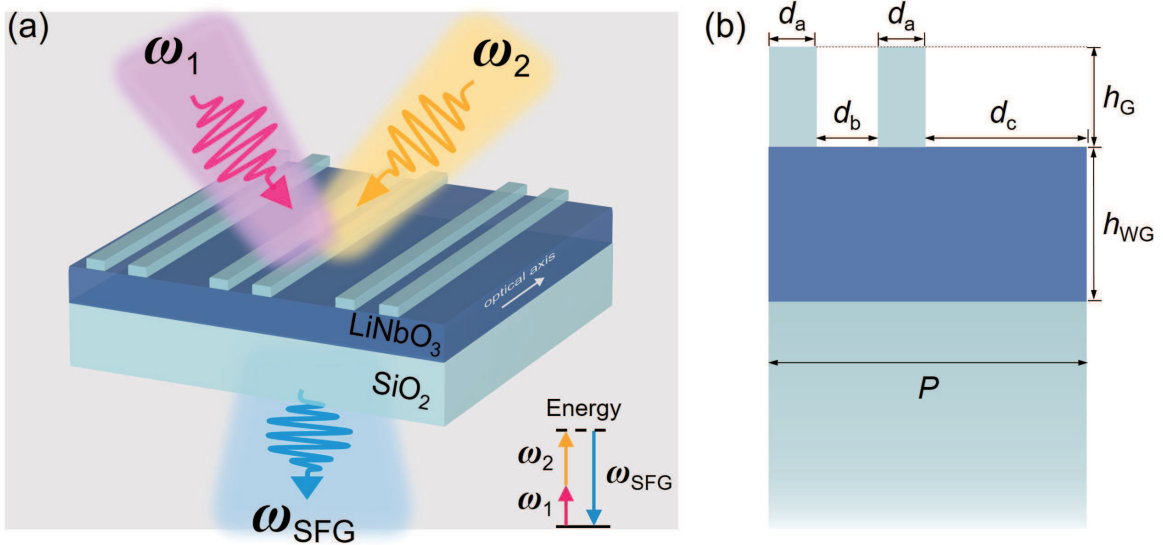}
	\caption{\label{fig1} (a) The schematic illustration of SFG process from the resonant grating waveguide structure composed of a SiO$_{2}$ four-part periodic grating, an etchless LiNbO$_{3}$ waveguide layer, and a SiO$_{2}$ substrate, from top to bottom. The inset shows the energy diagram of SFG process. (b) The cross section of the unit cell. The geometries are defined by the period $P=700$ nm, the widths of grating $d_{a}=0.15P$, $d_{\text{b}}=d_{0}-\Delta d$, and $d_{\text{c}}=d_{0}+\Delta d$, with $d_{0}=0.35P$, the heights of grating and waveguide $h_{\text{G}}=250$ nm and $h_{\text{WG}}=340$ nm, respectively.}
\end{figure*}

\begin{figure}[htbp]
	\centering
	\includegraphics% Here is how to import EPS art
	[scale=0.40]{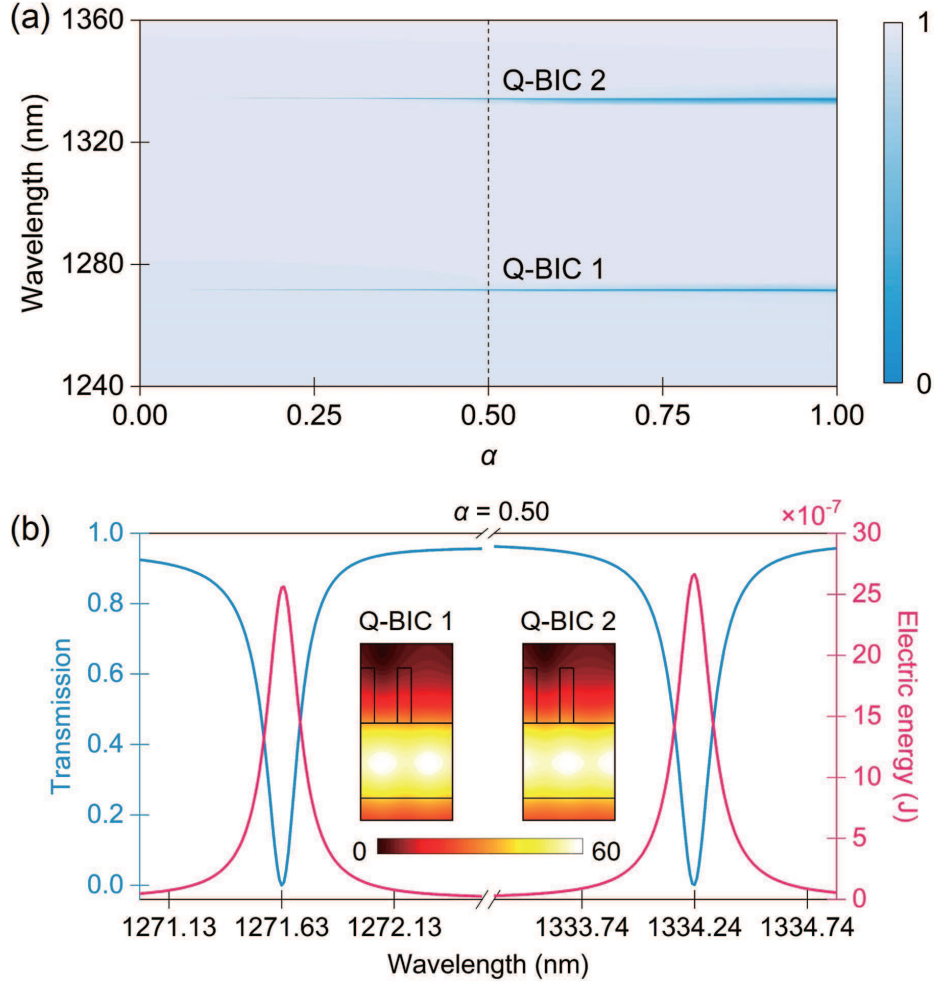}
	\caption{\label{fig2} (a) The evolution of transmission spectra of the resonant grating waveguide structure as a function of the grating geometry parameter $\alpha$ under oblique incidence $\theta=3^{\circ}$. (b) The transmission spectrum and electric energy inside the LiNbO$_{3}$ waveguide layer around the dual quasi-BICs with $\alpha=0.5$. The inset shows electric field distributions at the resonance wavelengths.}
\end{figure}

We start from the linear optical responses of the resonant grating waveguide structure. In the numerical calculation via finite element method software COMSOL Multiphysics, a two-dimensional electromagnetic model is adopted to simulate one unit cell with periodical boundary conditions applied in the $x$ direction and perfectly matched layers used in the $y$ direction. A transverse electric (TE) polarized plane wave (electric field is parallel to the $z$ direction) is incident at an oblique angle $\theta$. The refractive index of LiNbO$_{3}$ is extracted from the experimental measurement\cite{Zelmon1997}. To utilize the largest nonlinear tensor component of LiNbO$_{3}$ (which will be addressed in the nonlinear optical responses later), we align its crystallographic optical axis parallel to the incident polarization. Figure 2(a) presents a visual insight on the evolution of BICs in the grating waveguide structure under a fixed oblique angle $\theta=3^{\circ}$. When the grating geometry parameter $\alpha=0$, the dual-BICs, which in fact possess infinitely high $Q$ factors and vanishing resonant widths, are both hidden in the high-transmittance background; when $\alpha$ gradually increases to 1, the structural perturbations in the grating waveguide structure enable the coupling between the $\pm1^{\text{st}}$ order evanescent diffraction field and the leaky guided mode, and the genuine BICs transform to the quasi-BICs, corresponding to the excitation of two high-$Q$ guided mode resonances at around 1271 nm and 1334 nm, labeled by Q-BIC 1 and Q-BIC 2, respectively. The physical mechanism of the formation of dual BICs can be found from Sec. I in the Supplemental Material\cite{SM} and Refs. \cite{Wu2021, Liu2013, Yariv1984, Sang2011, Wu2019} therein. These BICs have also been termed as symmetry-unprotected BICs or Brillouin zone folding driven BICs in some metasurface systems\cite{He2018, Sun2023, Wang2023}. It is worth mentioning that the resonant wavelengths here are quite stable against the increase of $\alpha$ due to the constant filling ratio of SiO$_{2}$, in sharp contrast with the most well-known symmetry-protected quasi-BICs whose excitation requires the in-plane/out-of-plane symmetry breaking which always results in the resonant wavelength shifting and linewidth broadening at the same time\cite{Xu2019, Koshelev2019, Liu2019, Anthur2020, Gandolfi2021, Zograf2022, Xiao2022, Xiao2022a}. Figure 2(b) exemplifies the results with $\alpha=0.5$. The dual quasi-BICs can be clearly observed as two resonance dips located at 1271.63 nm (Q-BIC 1) and 1334.74 nm (Q-BIC 2). We derive $Q$ factors of the resonances by fitting the simulated transmission spectrum with the Fano formula expressed by $T=|a_{1}+ia_{2}+\frac{b}{\omega-\omega_{0}+i\gamma}|^{2}$, where $a_{1}$, $a_{2}$, and $b$ are real numbers, $\omega_{0}$ is the resonance frequency, and $\gamma$ is the leakage rate\cite{Xu2019, Koshelev2019, Wang2020a, Liu2023, Li2023, Li2023a}. Then the $Q$ factors (defined with $\omega_{0}/2\gamma$) are estimated as 7300 (Q-BIC 1) and 7211 (Q-BIC 2), respectively. Empowered by the guided mode resonances, the electric fields at both quasi-BICs are strongly confined inside the LiNbO$_{3}$ waveguide layer and significantly enhanced by a factor of about 60 compared with the incident electric field. Since SFG is a second-order process with a predominant volume contribution that responds mainly to the electric component of the incident plane wave\cite{Boyd2020}, we further analyze the electric energy stored inside the waveguide layer. The two maxima emerge exactly at the resonance wavelengths, highlighting the great potentials of the quasi-BICs in the enhanced excitation power for the SFG process.

\section{\label{sec3}Dual resonantly enhanced sum-frequency generation} 

Then we move forward to the nonlinear SFG process in the designed structure. It can be accurately calculated using coupled electromagnetic models in consecutive steps with the undepleted pump approximation: the first and second electromagnetic models are simulated at the input wavelengths to retrieve the local field distributions and compute the nonlinear polarization induced inside the nonlinear material LiNbO$_{3}$; and then this polarization term is employed as the only source for the third electromagnetic model at the sum-frequency wavelength to obtain the generated photon power flux propagated to the substrate. Considering that LiNbO$_{3}$ is a negative uniaxial birefringent crystal and belongs to point group 3m, the non-zero values of its anisotropic second-order nonlinear susceptibilities are $d_{31}=5.95$ pm/V, $d_{22}=3.07$ pm/V, and $d_{33}=34.4$ pm/V\cite{Jin2021}. To take advantage of the largest nonlinear coefficient $d_{33}$ of LiNbO$_{3}$, the input beams are linearly polarized along the same direction of optical axis, i.e. $z$-axis. Then the anisotropic nonlinear polarization of LiNbO$_{3}$ is written as
\begin{eqnarray}
	P_{x}^{\text{NL}}&=&2\varepsilon_{0}[d_{31}(E_{1x}E_{2z}+E_{1z}E_{2x})-d_{22}(E_{1y}E_{2x}+E_{1x}E_{2y})],\\
	P_{y}^{\text{NL}}&=&2\varepsilon_{0}[d_{31}(E_{1y}E_{2z}+E_{1z}E_{2y})+d_{22}(E_{1y}E_{2y}-E_{1x}E_{2x})],\\ 
	P_{z}^{\text{NL}}&=&2\varepsilon_{0}[d_{33}E_{1z}E_{2z}+d_{31}(E_{1y}E_{2y}+E_{1x}E_{2x})],\label{eq2}
\end{eqnarray}
where the subscripts $x$, $y$, and $z$ denote the corresponding components of the induced nonlinear polarization and electric field along different axes, and 1, 2 represent the first two electromagnetic models at the input wavelengths, respectively. In the linear simulations 1 and 2, the grating geometry parameter is set as $\alpha=0.5$, and the TE polarized plane waves are incident at $\lambda_{1}$ and $\lambda_{2}$ with an oblique angle $\theta=3^{\circ}$, which keep the same as in Fig. 2(b), and the input intensity is 100 MW/cm$^{2}$. Then in the nonlinear simulation, the SFG power is calculated in the third electromagnetic model by integrating the Poynting vector over the output plane on the transmission side. 

The SFG process strength is characterized by the conversion efficiency defined with $\eta_{\text{SFG}}=P_{\text{SFG}}/(P_{1}+P_{2})$, where $P_{\text{SFG}}$ is the above-mentioned the output power of the generated SFG signal, and $P_{1}$ and $P_{2}$ are literally the power of the two input beams. As depicted in Fig. 3(a), SFG process shows strong dependence on the input wavelengths. The maximum conversion efficiency can reach as high as $\eta_{\text{SFG}}=3.66\times10^{-2}$ in the case of $\lambda_{1}=1271.63$ nm and $\lambda_{2}=1334.24$ nm, which results in a generated SFG signal at the visible wavelength of 651.09 nm. In addition, an important feature captured here is the giant enhancement of the SFG efficiency limited in a narrow spectral range, with a small full-width at half-maximum less than 0.2 nm. It is the direct consequence of the ultrahigh $Q$-factor resonances with extremely narrow resonant widths derived from the quasi-BIC features as depicted in Fig. 2. To gain a clear insight of resonant enhancement effect driven by the dual quasi-BICs, Figs. 3(c) and (d) exemplify the results of the simulated SFG conversion efficiency around the resonances. With the wavelength of one input beam fixed at $\lambda_{2}=1334.24$ nm (Q-BIC 2), $\eta_{\text{SFG}}$ reaches its peak when the wavelength of the other input beam locates at $\lambda_{1}=1271.63$ nm (Q-BIC 1), and vice versa. The maximum enhancement of the SFG process appears only with the simultaneous excitation of the two quasi-BICs at their respective wavelengths. Due to the strongly enhanced local electric fields inside the LiNbO$_{3}$ waveguide layer at these two quasi-BICs wavelengths, as shown in the inset of Fig. 2(b), $\eta_{\text{SFG}}$ here is increased by a factor of $10^{5}$ compared with the LiNbO$_{3}$ films of the same thickness. On the other hand, $\eta_{\text{SFG}}$ depends on the input intensity as well. The linear dependence of $\eta_{\text{SFG}}$ on the input intensity can be derived from Fig. 3(b), implying that higher conversion efficiency up to the order of $10^{-1}$ can be realized by further increasing the input intensity to the order of GW/cm$^{2}$. Such high conversion efficiency is comparable or even further improved to previously theoretical results with quasi-BICs enhanced second-order nonlinear processes\cite{Carletti2018, Yang2020, Ning2021, Huang2021, Kang2021, Parry2021, Zheng2022, Ge2023}. 

\begin{figure}[htbp]
	\centering
	\includegraphics% Here is how to import EPS art
	[scale=0.36]{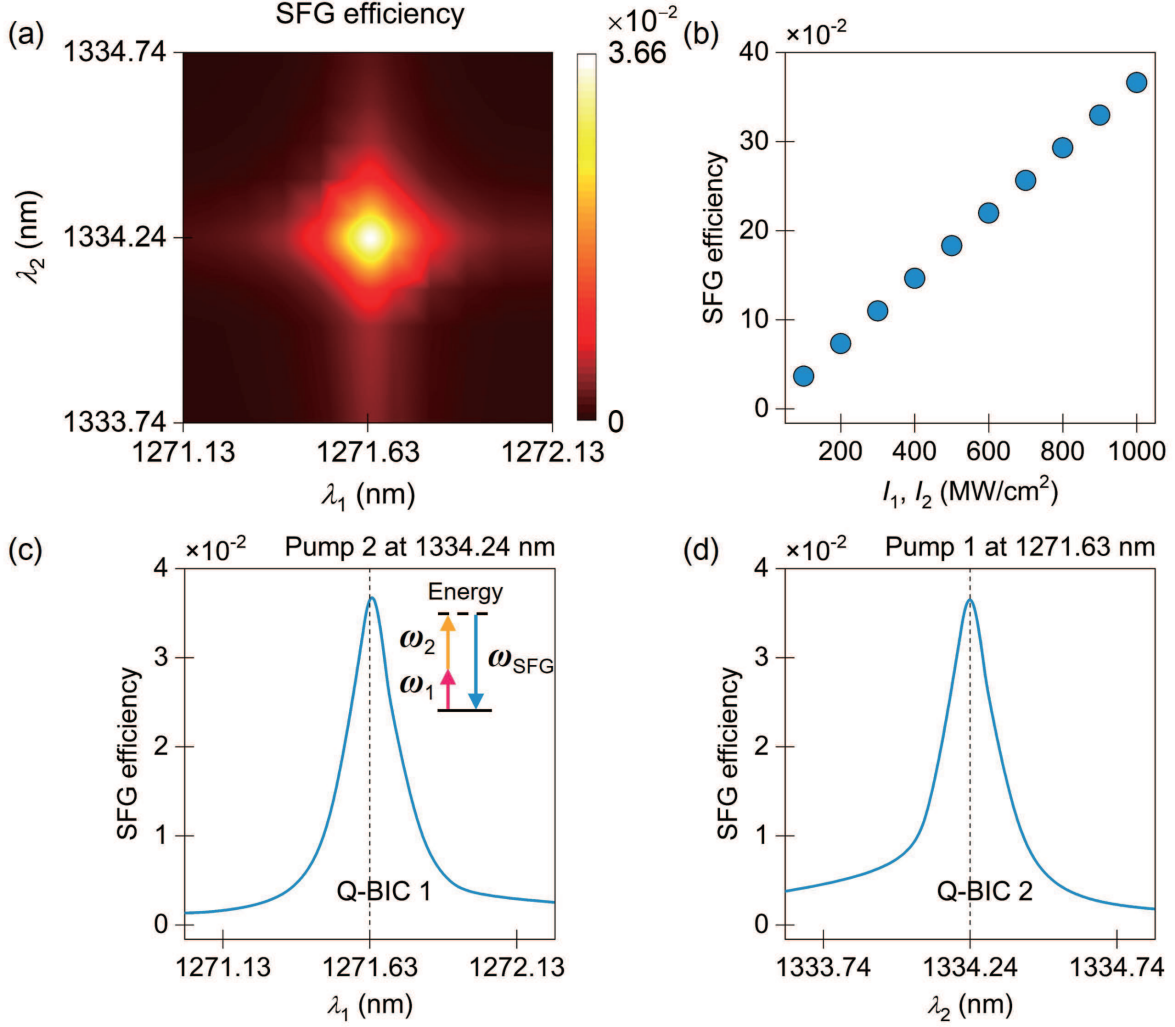}
	\caption{\label{fig3} (a) The SFG conversion efficiency as a function of the two input wavelengths around the quasi-BICs. (b) The SFG conversion efficiency as a function of the input intensity. (c) and (d) The SFG conversion efficiency as a function of the input beam 1(2) wavelength when the input beam 2(1) wavelength is kept constant at quasi-BIC 2(1).}
\end{figure}

\begin{figure}[htbp]
	\centering
	\includegraphics% Here is how to import EPS art
	[scale=0.36]{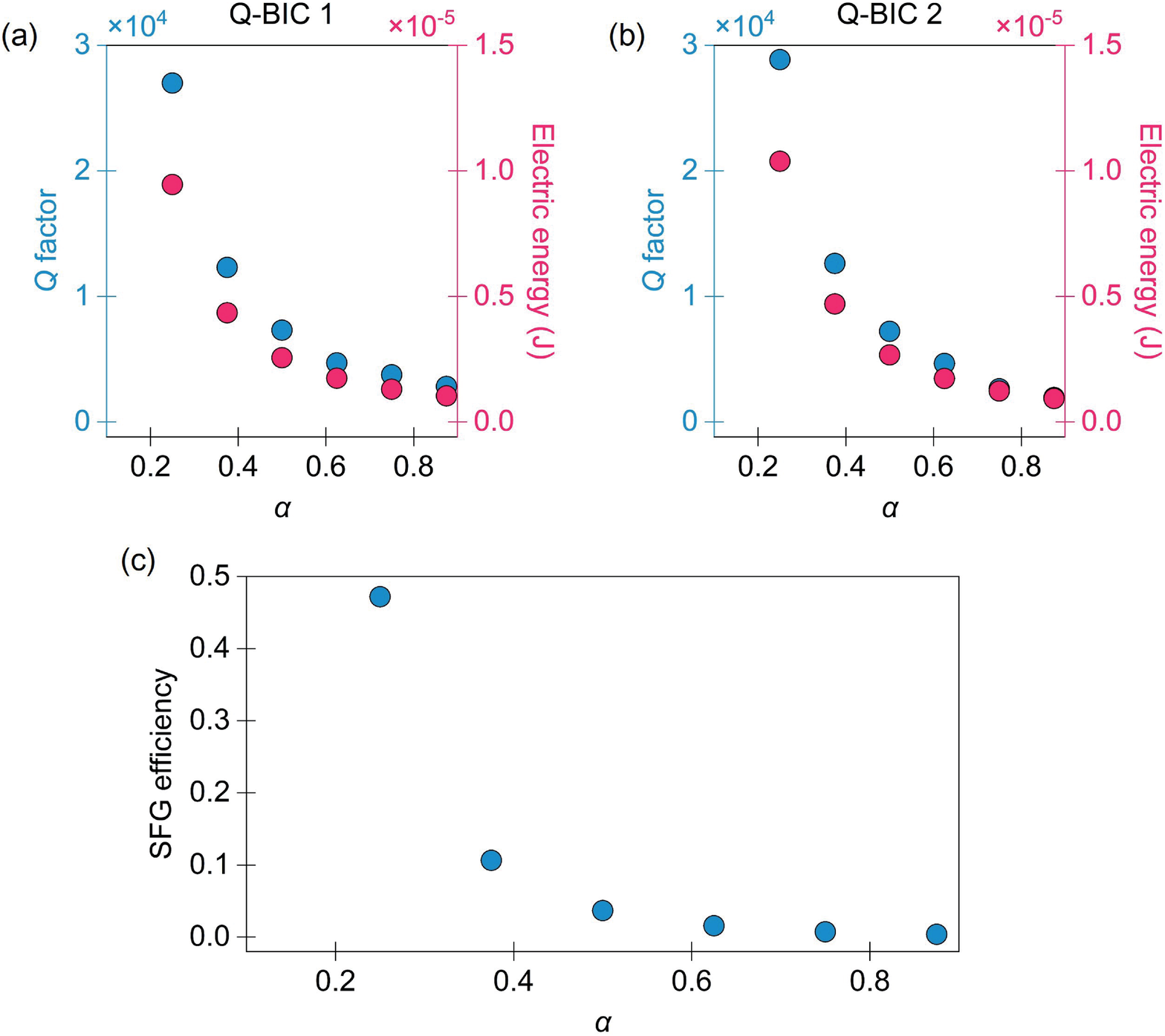}
	\caption{\label{fig4} (a) and (b) The $Q$ factor and electric energy insider the LiNbO$_{3}$ waveguide layer as a function of the grating geometry parameter $\alpha$ at the quasi-BICs. (c) The SFG conversion efficiency as a function of $\alpha$ at the quasi-BICs.}
\end{figure}

Next, we explore the effect of $Q$ factor on the SFG conversion efficiency in the designed structure. The grating geometry parameter $\alpha$ can be utilized to engineer the radiative loss rate of the system and directly affect the $Q$ factor of the quasi-BICs. As shown in Figs. 4(a) and (b), the $Q$ factors of Q-BIC 1 and Q-BIC 2 follow the inverse quadratic dependence on the parameter $\alpha$, i.e., $Q\propto Q_{0}\alpha^{-2}$, where $Q_{0}$, with fitting values of 1644.13 and 1825.56 for Q-BIC 1 and Q-BIC 2, respectively, are constants determined by the structure design\cite{Koshelev2018, Li2019, Kutuzova2023}. When $\alpha$ is close to 0, both $Q$ factors approach the infinite corresponding to the genuine BICs. When $\alpha$ gradually increases to 1, the $Q$ factors dramatically decline. Since the $Q$ factor is related directly to the capability of electric field confinement, the electric energy inside the LiNbO$_{3}$ waveguide layer shows a similar decline tendency. Considering that the SFG process is essentially the combined results of the two input beams, its efficiency is jointly determined by the $Q$ factors of the dual quasi-BICs, and approximately scales as the product $Q_{1}\times Q_{2}$, where $Q_{1}$ and $Q_{2}$ are the $Q$ factors of Q-BIC 1 and Q-BIC 2, respectively. As a result, the conversion efficiency of the SFG process shows a steeper decline as $\alpha$ increases, as illustrated in Fig. 4(c). It can be tuned over two orders of magnitude ranging from $10^{-1}$ to $10^{-3}$ via the varying $Q$ factor, but remains locating at a quite stable SFG wavelength around 651 nm since the resonant wavelengths of the quasi-BICs are robust against the increase of $\alpha$. It is also noted that the sensitivity of the $Q$ factor to the geometry parameter $\alpha$ brings the possibility to obtain higher conversion efficiency of the SFG process considering that $Q$ factor provides the upper boundary for the enhancement of nonlinear light-matter interactions. The conversion efficiency of $10^{-2}$ as we claimed before is the result of $Q$ factors around 7000 by the conservative choice of $\alpha=0.5$. It is a trade off after fully taking the fabrication tolerance into consideration. At $\alpha=0.5$, the width difference between the second and fourth parts of the top grating is larger than 200 nm, which is well within the reach of current fabrication technique\cite{Koshelev2019, Liu2019, Wang2021, Ma2021, Zograf2022, Chen2023a, Wu2023, Maggiolini2023}. 

We further calculate the dependence of the SFG efficiency on the polarization angle of the input beams, as shown in Fig. 5. The wavelengths of the input beams are fixed at 1271.63 nm and 1334.74 nm, respectively, and the oblique incident angle is still fixed as $\theta=3^{\circ}$, in which case the dual quasi-BICs resonantly enhanced SFG process has been obtained. At first, we rotate the polarization of input beam 1 over $0^{\circ}-360^{\circ}$ while maintaining fixed polarization of the input beam 2 along the $z$-axis. It can be observed that the maximum SFG efficiency is obtained when the input beam 1 is polarized at $0^{\circ}$ and $180^{\circ}$, namely, along the $z$-axis; while the minimum is achieved when the input beam 1 is polarized at $90^{\circ}$ and $270^{\circ}$, namely, along the $x$-axis (blue solid line). Then we rotate the polarization of input beam 2 with fixed polarization of the input beam 1 along the $z$-axis, and observe the same polarization dependence (red dashed line). In both cases, the maximum SFG efficiency is obtained for the concurrent TE polarization incidence of the input beams with their electric fields along $z$-axis. The observed polarization dependence of SFG agrees well with the linear simulations where the dual quasi-BICs are both preferably excited in the resonant grating waveguide structure under the $z$-axis polarization. Owing to the polarization sensitivity of the designed one-dimensional structure, the SFG efficiency can be continuously tuned between its maximum and minimum via simply choosing the input beam polarization combination. This tuning strategy has once been implemented in nonlinear metasurfaces\cite{Xiao2020, Carletti2021}. In addition, the anisotropic nature of second-order nonlinear susceptibilities of LiNbO$_{3}$ also affects the polarization properties of SFG. The input beams under $z$-axis polarization, which is parallel to the LiNbO$_{3}$ optical axis, utilize the largest nonlinear tensor component $d_{33}$, leading to higher maximum of the conversion efficiency.

\begin{figure}[htbp]
	\centering
	\includegraphics% Here is how to import EPS art
	[scale=0.24]{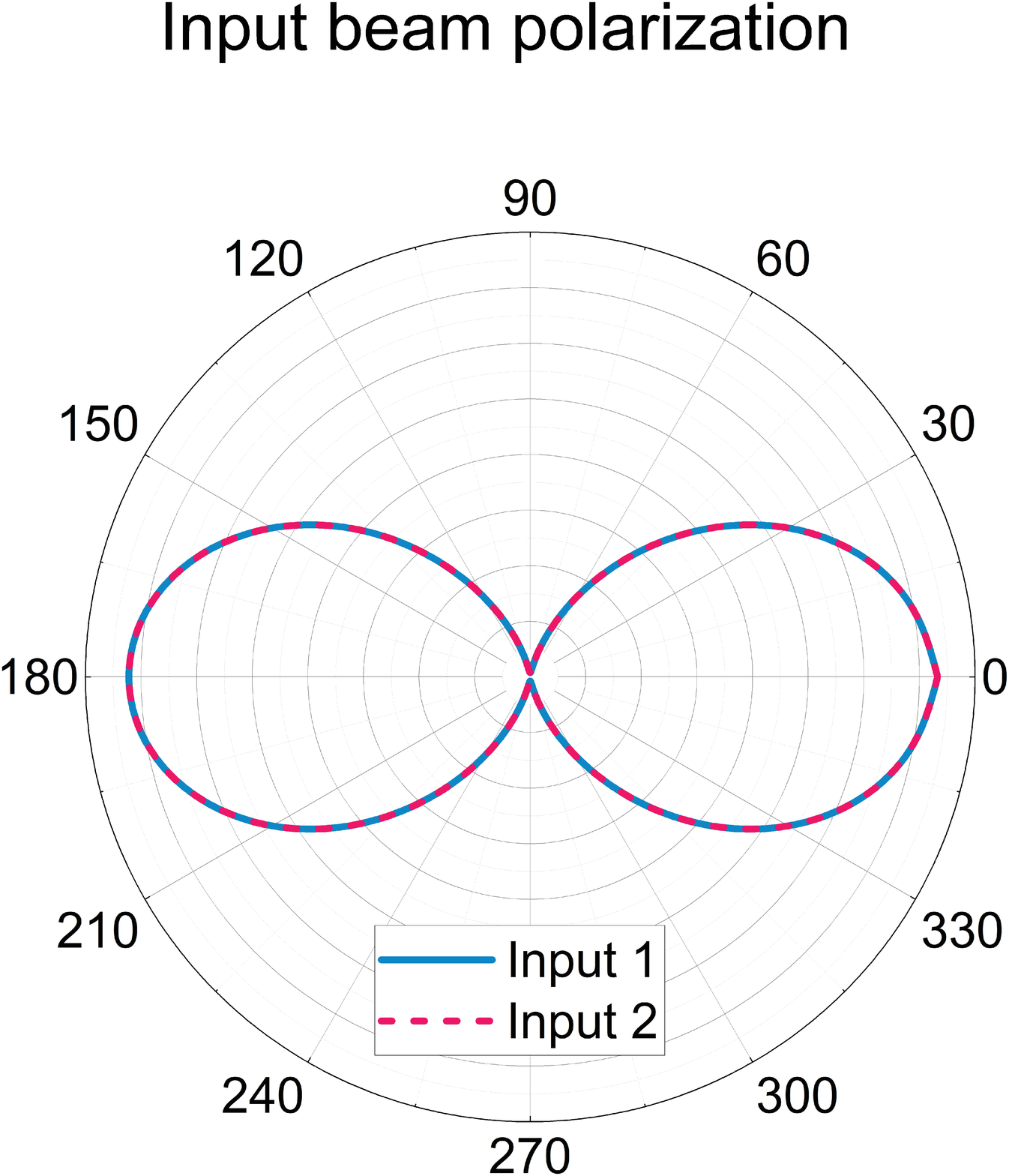}
	\caption{\label{fig5} The polar plots for the SFG conversion efficiency as a function of the polarization angle of the input beam 1(2) when maintaining fixed polarization of the input beam 2(1) along the $z$-axis. In both plots, $0^{\circ}$ and $90^{\circ}$ represent polarizations along the $z$- and $x$-axis, respectively. The SFG efficiencies are normalized to the maximum.}
\end{figure}

\begin{figure}[htbp]
	\centering
	\includegraphics% Here is how to import EPS art
	[scale=0.40]{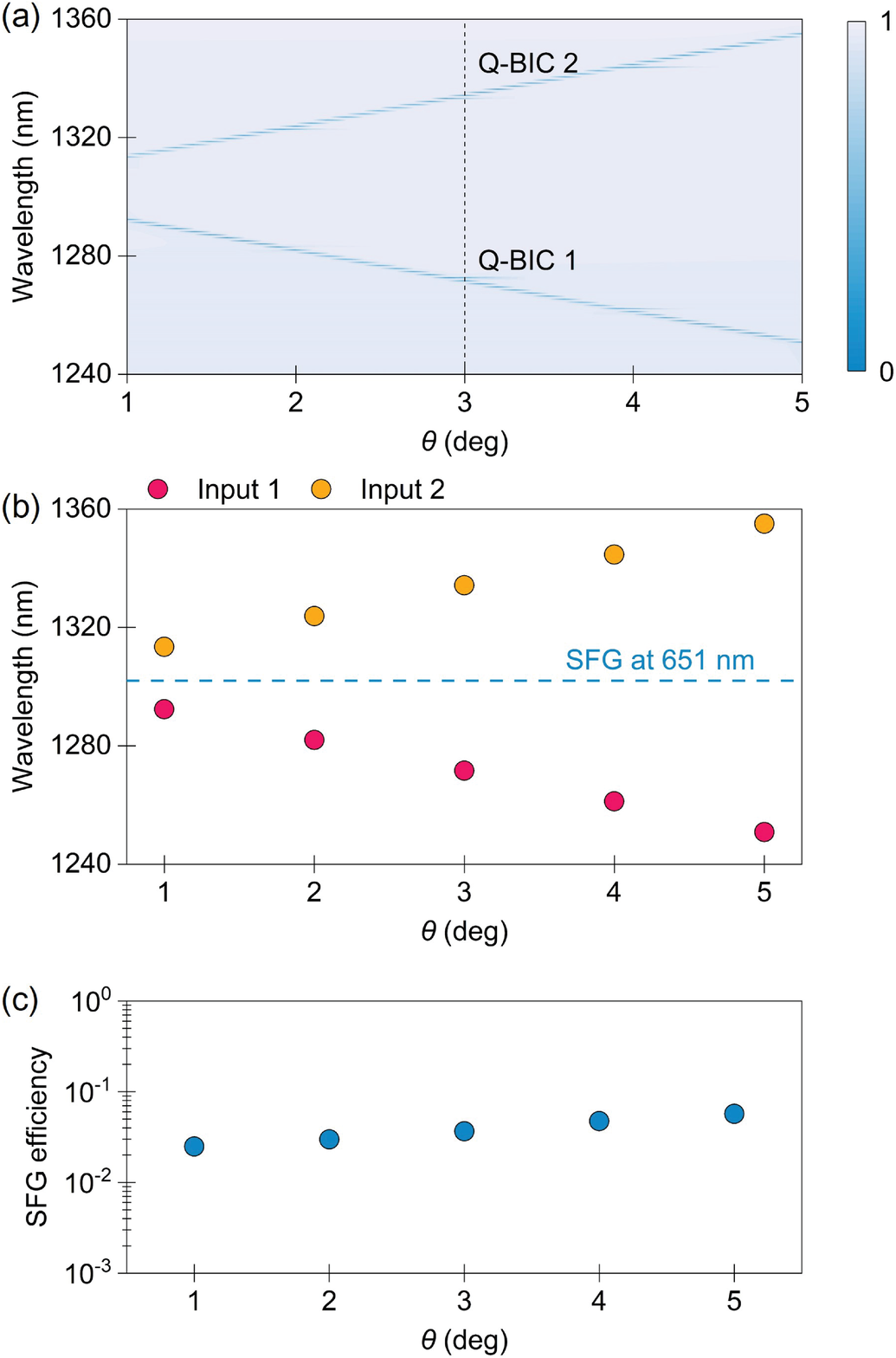}
	\caption{\label{fig6} (a) The evolution of transmission spectra of the resonant grating waveguide structure as a function of the incident angle $\theta$. (b) The fixed SFG wavelength of 651 nm when the wavelengths of the two input beams are jointly tuned by selecting $\theta$. (c) The SFG conversion efficiency as a function of $\theta$ at the corresponding quasi-BICs.}
\end{figure}

Finally, we analyze the impact of the incident angle on the SFG process. As seen from Sec. II in the Supplemental Material\cite{SM}, the oblique incidence that eliminates the degenerate of the $\pm1^{\text{st}}$ order resonant guided modes is the key for realizing the dual quasi-BICs. It can be seen in Fig. 6(a), as the incident angle increases from $1^{\circ}$ to $5^{\circ}$, the Q-BIC 1 shows a blueshift toward shorter wavelength regimes and the Q-BIC 2 shows an opposite redshift. The larger incident angle gives rise to larger difference between the two resonant wavelengths. This angle-dependent behavior is similar to that of the guided mode resonances in the conventional two-part periodic grating waveguide structure\cite{Liu1998}, but the quasi-BICs provide much higher $Q$ factors and stronger local electric field confinement for nonlinear light-matter interactions. Intriguingly, despite that the dual quasi-BICs show the angle-dependence, their SFG wavelength keeps unchanged, which brings an important striking feature to the designed structure in robustness. The narrowband SFG signal can be generated at a fixed wavelength with tunable wavelengths of input beams. As illustrated in Fig. 6(b), the SFG emission can be maintained at the fixed wavelength of 651 nm, while the wavelengths of the two input beams are jointly tuned over a broad spectral range by selecting incident angles. For example, under oblique incidence $\theta=5^{\circ}$, the two input beams at the quasi-BIC wavelengths 1250.89 nm and 1355.06 nm would also generate the dual resonantly enhanced SFG signal at around 651 nm that is exactly the SFG wavelength in the case of $3^{\circ}$ incidence. Even more interesting, Fig. 6(c) further verifies that the SFG efficiency remains considerably high for different incident angles (and thus different input wavelengths) as long as the corresponding dual quasi-BICs are excited. Such high conversion efficiency is the result of high-$Q$ factors which can be easily captured from the linear transmission spectra in Fig. 6(a). Therefore, it can also be envisaged that the dual quasi-BICs that support the high-efficiency SFG process here would enhance the reverse SPDC process as well. In the designed structure, the pump photons (at the SFG wavelength) normally impinge from the substrate side, split into signal and idler photons (at quasi-BIC wavelengths) in the waveguide layer via SPDC, and the generated photon pairs finally pass through the grating and get emitted to air at an oblique angle. These photon pairs would be entangled with angle-correlated and frequency non-degenerate properties, which would be beneficial for quantum sources. Previous works have revealed the potential of dual quasi-BICs in boosting the SPDC process in high-index nonlinear dielectric materials\cite{Parry2021, Mazzanti2022, Santiago-Cruz2022, Son2023}. The angle-dependent dual quasi-BICs here may facilitate the efficient generation of complex quantum states, and in particular, enable the flexible manipulation of the quantum features of entangled photons. 

%It can be predicted using either the quantum calculation or the quantum-classical correspondence between SPDC and SFG\cite{Poddubny2016}, however, it would be beyond the scope of the present study so we prefer to stop here.

\section{\label{sec4}Conclusions}

In conclusion, we have demonstrated high-efficiency SFG emission from etchless LiNbO$_{3}$ empowered by dual quasi-BICs in a resonant grating waveguide structure. Under oblique incidence, the designed structure can support two ultrahigh-$Q$ guided mode resonances corresponding to the excitation of the quasi-BICs, which provide strong enhancement of local electric fields for nonlinear frequency conversion. A narrowband visible SFG signal is generated with conversion efficiency substantially increased by five orders of magnitude compared with the LiNbO$_{3}$ films of the same thickness. The conversion efficiency of SFG can be further tuned over two orders of magnitude with varying Q factors via adjusting the grating geometry parameter. Owing to the polarization sensitivity of the dual quasi-BICs and the anisotropic nature of second-order nonlinear susceptibilities of LiNbO$_{3}$, the maximum SFG efficiency is obtained for the concurrent TE polarization incidence of the input beams with their electric fields along $z$-axis, which, in turn, provides continuously tunable emission properties by simply choosing the input polarization combination. We also find the robustness of SFG signal that it can be generated at a fixed wavelength with tunable wavelengths of input beams, which may be in particular preferable for its reverse SPDC process, i.e., the generation of angle-correlated and frequency non-degenerate entangled photon pairs. The high conversion efficiency and engineered nonlinearity from such resonant grating waveguide structure offer the possibility of designing high-performance on-chip nonlinear devices such as tunable laser sources, ultrathin optical correlators, and entangled photon pairs generators. Finally, it is worth pointing out that the method and structure here are general, and can thus be extended for the enhancement of other nonlinear effects at alternative wavelengths via the physical scalability. The first example that comes to mind may be the harmonic generation and frequency mixing in two-dimensional materials\cite{Yuan2019, Yao2019, Kim2020, Bernhardt2020, Loechner2020, Liu2021}.

\begin{acknowledgments}	
	
This work is supported by the National Natural Science Foundation of China (Grants No. 11947065, No. 12104105, and No. 12264028), the Natural Science Foundation of Jiangxi Province (Grant No. 20202BAB211007), the Guangdong Basic and Applied Basic Research Foundation (Grant No. 2023A1515011024), the Interdisciplinary Innovation Fund of Nanchang University (Grant No. 2019-9166-27060003), the Open Project of Shandong Provincial Key Laboratory of Optics and Photonic Devices (Grant No. K202102), the Start-up Funding of Guangdong Polytechnic Normal University (Grant No. 2021SDKYA033), and the China Scholarship Council (Grant No. 202008420045). The authors would like to thank Prof. Tingyin Ning, Prof. Lujun Huang, and Prof. Tianjing Guo for beneficial discussions on the BIC physics and nonlinear numerical simulations.

S. F. and T. L. contributed equally to this work.

\end{acknowledgments}

%\bibliography{Ref}% Produces the bibliography via BibTeX.

%apsrev4-2.bst 2019-01-14 (MD) hand-edited version of apsrev4-1.bst
%Control: key (0)
%Control: author (8) initials jnrlst
%Control: editor formatted (1) identically to author
%Control: production of article title (0) allowed
%Control: page (0) single
%Control: year (1) truncated
%Control: production of eprint (0) enabled
%

\end{document}